\documentclass[prd,preprint,superscriptaddress,showpacs,nofootinbib,%
tightenlines,floatfix]{revtex4}

\usepackage{amsmath,amssymb,slashed,epsfig,subfigure,float,multirow}

\def\bra#1{\ensuremath{\langle{#1}\vert}}
\def\ket#1{\ensuremath{\vert{#1}\rangle}}

\newcommand{ \slashchar }[1]{\setbox0=\hbox{$#1$}   
   \dimen0=\wd0                                     
   \setbox1=\hbox{/} \dimen1=\wd1                   
   \ifdim\dimen0>\dimen1                            
      \rlap{\hbox to \dimen0{\hfil/\hfil}}          
      #1                                            
   \else                                            
      \rlap{\hbox to \dimen1{\hfil$#1$\hfil}}       
      /                                             
   \fi}

\begin{document}
\title{Electromagnetic form factors of the nucleon\\ in effective field theory}
\author{T.~Bauer}
\affiliation{Institut f\"ur Kernphysik, Johannes
Gutenberg-Universit\"at, D-55099 Mainz, Germany}
\author{J.~C.~Bernauer}
\affiliation{Institut f\"ur Kernphysik, Johannes
Gutenberg-Universit\"at, D-55099 Mainz, Germany}
\affiliation{Present address: Laboratory for Nuclear Science, MIT, Cambridge, MA 02139, USA}
\author{S.~Scherer}
\affiliation{Institut f\"ur Kernphysik, Johannes
Gutenberg-Universit\"at, D-55099 Mainz, Germany}
\date{September 18, 2012}

\begin{abstract}
   We calculate the electromagnetic form factors of the nucleon to third
chiral order in manifestly Lorentz-invariant effective field theory.
   The $\rho$ and $\omega$ mesons as well as the $\Delta(1232)$ resonance
are included as explicit dynamical degrees of freedom.
   To obtain a self-consistent theory with respect to constraints we consider
the proper relations among the couplings of the effective Lagrangian.
   For the purpose of generating a systematic power counting, the extended
on-mass-shell renormalization scheme is applied in combination with the
small-scale expansion.
   The results for the electric and magnetic Sachs form factors are analyzed
in terms of experimental data and compared to previous findings in
the framework of chiral perturbation theory.
   The pion-mass dependence of the form factors is briefly discussed.
\end{abstract}

\maketitle
\section{Introduction}

   Electromagnetic form factors parameterize the single-nucleon matrix
element of the electromagnetic current operator and provide important
information about the structure and composition of the nucleon
(see, e.g., Refs.~\cite{Gao:2003ag,Perdrisat:2006hj,Drechsel:2007sq}
for an overview).
   Furthermore, they are important input to high-precision tests of quantum
electrodynamics as well as the Standard Model of particle physics.
   In the space-like region, the proton form factors have been measured with great
accuracy over a wide range of momentum transfer in experiments on
elastic electron-nucleon scattering.
   Neutron form factors are not as well known since they have to be extracted
from scattering experiments with deuterium or $^3 \textrm{He}$.
   Despite the wealth of the available data there are still open issues
such as the value of the proton charge radius determined from the Lamb shift in muonic hydrogen on the one hand
\cite{Mohr:2012tt} and electronic hydrogen Lamb shift measurements and elastic electron-proton scattering on the
other hand \cite{Beringer:1900zz}.
   A new generation of precision measurements of electromagnetic
form factors at low momentum transfer has been and is presently performed at
the Mainz Microtron MAMI \cite{Bernauer:2008zz,Bernauer:2010wm}.

   Chiral perturbation theory (ChPT)
\cite{Weinberg:1978kz,Gasser:1983yg,Gasser:1987rb} is the effective
field theory (EFT) of quantum chromodynamics in the low-energy
domain (for an introduction and review see, e.g.,
Refs.~\cite{Scherer:2009bt,Scherer:2012zzd}).
   The first form factor calculation was performed in the early relativistic
approach \cite{Gasser:1987rb}, in which, however, the power counting of low-energy
dimensions was still an open issue due to the additional heavy-mass scale introduced by
the nucleon.
   Later on, the problem of setting up a consistent power counting in EFT with heavy
degrees of freedom was handled by employing the heavy-baryon approach
\cite{Jenkins:1990jv,Bernard:1992qa} and, more recently, by choosing suitable
renormalization prescriptions in a manifestly Lorentz-invariant framework
\cite{Tang:1996ca,Becher:1999he,Gegelia:1999gf,Fuchs:2003qc}.
   Within the heavy-baryon approach, calculations of the form factors were performed in
Refs.~\cite{Bernard:1992qa,Fearing:1997dp} and, including the $\Delta(1232)$ resonance
in terms of the small-scale expansion \cite{Hemmert:1997ye}, in Ref.~\cite{Bernard:1998gv}.
   Applying different renormalization schemes, form factor calculations
have been performed within manifestly Lorentz-invariant baryon ChPT
up to $\mathcal{O}(q^4)$ \cite{Kubis:2000zd,Fuchs:2003ir}
and to $\mathcal{O}(q^3)$ including the leading-order
corrections due to the $\Delta$ resonance \cite{Ledwig:2011cx}.
   In general, such calculations describe the experimental data only for a small range
of momentum transfer ($Q^2\leq 0.1~ \textrm{GeV}^2$).
   On the other hand, the $\rho$, $\omega$, and $\phi$ mesons were included
dynamically in the effective Lagrangians of
Refs.~\cite{Kubis:2000zd,Schindler:2005ke}.
   A systematic re-summation of higher-order terms which, in an ordinary chiral expansion,
would contribute at higher orders beyond ${\cal O}(q^4)$,
results in an improved description of the data even for higher values
of $Q^2$, as expected on phenomenological grounds.
   The re-organization proceeds according to well-defined rules \cite{FSGS03} so that a
controlled, order-by-order calculation of corrections is made possible.

   A covariant formalism for massive vector fields involves Lagrangians with constraints,
because one typically introduces unphysical degrees of freedom \cite{weinbergQTFI}.
   In comparison with Refs.~\cite{Kubis:2000zd,Schindler:2005ke}, the present paper considers the
conditions on the form of the Lagrangian imposed by the demand for a
self-consistent theory in terms of constraints
\cite{Djukanovic:2004mm,Djukanovic:2010tb}.
   We use the extended on-mass-shell (EOMS) scheme \cite{Fuchs:2003qc} to generate a
systematic power counting in the presence of heavy degrees of
freedom \cite{FSGS03}.
   As a result, we obtain an effective Lagrangian which is renormalizable
in the sense of effective field theory \cite{weinbergQTFIb} and which is
consistent with the constraints order by order.

   Because of its close proximity to the ground state and its strong
coupling to the pion-nucleon-photon system we also include the
$\Delta(1232)$ resonance explicitly in our effective theory.
   The nucleon-$\Delta$ mass splitting is treated as an additional small parameter
(small-scale expansion \cite{Hemmert:1997ye}).
   Akin to the vector-meson case we respect the constraints on the possible
interaction terms to obtain a self-consistent theory describing the right number
of degrees of freedom \cite{Wies:2006rv,Hacker:2005fh}.

   This paper is organized as follows.
   In Sec.\ II the definitions of the Dirac and Pauli as well as the
Sachs form factors are given.
   We briefly discuss those elements of the most general effective
Lagrangian relevant for the subsequent calculation and state the
applied power-counting rules in Sec.\ III.
   In Sec.\ IV we discuss the fit of our results to experimental data.
   The final results for the Sachs form factors are presented and analyzed.
   Section V contains a short summary.

\section{Electromagnetic Form Factors of the Nucleon}

   Neglecting the contributions due to heavier quarks, the
electromagnetic current operator is given by
\begin{equation}
J^\mu(x)=\frac{2}{3}\bar{u}(x)\gamma^\mu
u(x)-\frac{1}{3}\bar{d}(x)\gamma^\mu
d(x)=\bar{q}(x)\left(\frac{1}{6}+\frac{\tau_3}{2}\right)\gamma^\mu
q(x),
\end{equation}
and the interaction with an external electromagnetic four-vector potential
$\mathcal{A}_\mu$ reads
\begin{equation}
\mathcal{L}_{\textrm{e.m.}}=-e J^\mu \mathcal{A}_\mu,
\end{equation}
where $e>0$ denotes the elementary charge.
   In the one-photon-exchange approximation, the electromagnetic form
factors are defined via the matrix element
\begin{align}
\bra{N(p_f)}J^{\mu}(0)\ket{N(p_i)}=
\bar{u}(p_f)\left[\gamma^{\mu}F_{1}^{N}(Q^2)+\frac{i\sigma^{\mu
\nu}q_{\nu}}{2m_p}F_{2}^{N}(Q^2)\right]u(p_i),\qquad
N=p,n,\label{formfactor}
\end{align}
where $m_p$ denotes the proton mass, $q=p_f-p_i$ is the four-momentum
transfer, and $Q^2\equiv-q^2\geq0$.
   The functions $F_1^N (Q^2)$ and $F_2^N (Q^2)$ are called Dirac and
Pauli form factors, respectively.
   At $Q^2=0$, the Dirac form factor takes the value of the electric
charge in units of the elementary charge and the Pauli form factor
takes the value of the anomalous magnetic moment in units of the
nuclear magneton:
\begin{align}
F_{1}^{p}(0)&=1, \qquad F_{2}^{p}(0)=1.793,\\
F_{1}^{n}(0)&=0,\qquad F_{2}^{n}(0)=-1.913.\label{magneticmoments}
\end{align}
   Our final results will be displayed in terms of the
electric and magnetic Sachs form factors \cite{Ernst:1960zza} since these are better
suited for the analysis of experimental data.
   The Sachs form factors are related to the Dirac and Pauli form factors as follows:
\begin{equation}
\label{sachsdef}
\begin{split}
G_E^N(Q^2) &= F_1^N(Q^2)-\frac{Q^2}{4m_p^2} F^N_2(Q^2),\\
G_M^N(Q^2) &= F_1^N(Q^2)+ F^N_2(Q^2).
\end{split}
\end{equation}
   Sometimes it is more convenient to work with the isoscalar and
isovector form factors defined as the sum and difference of the
proton and neutron form factors, respectively,
\begin{equation}
F_i^{(s)}=F_i^p+F_i^n,\quad F_i^{(v)}=F_i^p-F_i^n,\quad i=1,2.
\end{equation}
   The isoscalar and isovector Sachs form factors are defined accordingly.

\section{Effective Lagrangian and power counting}

\subsection{Non-resonant Lagrangian}
   The non-resonant part of the effective Lagrangian consists of a
purely mesonic part and a part describing the interaction of pions
and nucleons.
   From the mesonic sector only the lowest-order
Lagrangian ${\cal L}_{2}$, including the coupling to an external
electromagnetic four-vector potential $\mathcal{A}_\mu$ in terms of
the isovector field $v_\mu=-e \mathcal{A}_\mu \tau_3/2$, is needed
\cite{Gasser:1983yg},
\begin{equation}
\label{L2}
{\cal L}_{2}=\frac{F^2}{4}\textrm{Tr} \left(\partial_\mu U\partial^\mu U^\dagger\right)
+\frac{F^2 M^2}{4}\textrm{Tr}\left(U^\dagger+U\right)
+i\frac{F^2}{2}\textrm{Tr}\left[(\partial^\mu U U^\dagger+\partial^\mu U^\dagger U)v_\mu\right].
\end{equation}
   Here, $F$ denotes the pion-decay constant in the chiral limit,
$F_\pi=F\left[1+\mathcal{O}(\hat{m})\right]=92.2$ MeV, and $M^2=2
B\hat{m}$ is the squared pion mass at leading order in the
quark-mass expansion.
   In the isospin-symmetric limit $\hat m=m_u=m_d$,
and $B$ is related to the scalar singlet
quark condensate $\left\langle \bar{q}q\right\rangle_0$
in the chiral limit \cite{Gasser:1983yg,Colangelo:2001sp}.
   The pion fields are contained in the unimodular, unitary matrix $U$:
\begin{displaymath}
U(x)=u^2(x)=\exp\left(i\frac{\phi(x)}{F}\right),\quad
\phi=\phi_k\tau_k.
\end{displaymath}

   Collecting the proton and nucleon fields in the isospin doublet $\Psi$,
the lowest-order $\pi N$ Lagrangian is given by \cite{Gasser:1987rb}
\begin{equation}
{\cal L}_{\pi N}^{(1)}=\bar{\Psi}\left(i\slashed{D}-m+\frac{\texttt{g}_A}{2}\gamma^\mu
\gamma_5 u_\mu \right)\Psi,\label{LpiN}
\end{equation}
with
\begin{align*}
D_\mu \Psi &=\left(\partial_\mu+\Gamma_\mu-iv_\mu^{(s)}\right)\Psi,\\
\Gamma_\mu &= \frac{1}{2}\left[u^\dagger \partial_\mu u+u
\partial_\mu u^\dagger-i\left(u^\dagger v_\mu u +u v_\mu
u^\dagger\right)\right],
\\
u_\mu &=i\left[u^\dagger
\partial_\mu u-u \partial_\mu u^\dagger-i\left(u^\dagger v_\mu u -u
v_\mu u^\dagger\right)\right],
\end{align*}
where $ v_{\mu}^{(s)}=-e \mathcal{A}_\mu/2$
   In Eq.~(\ref{LpiN}), $m$ and $\texttt{g}_A$ denote the chiral limit
of the physical nucleon mass and the axial-vector coupling constant,
respectively.

   The complete Lagrangians at second and third order can be found in
Ref.~\cite{Ecker:1995rk}. We only display those terms needed for our
calculation,
\begin{equation}
\begin{split}
{\cal L}_{\pi N}^{(2)}&=
\bar{\Psi}\sigma^{\mu\nu}\left(\frac{c_6}{2}f^+_{\mu\nu}+\frac{c_7}{2}v^{(s)}_{\mu\nu}\right)\Psi
+ \cdots \:,\\
{\cal L}_{\pi N}^{(3)}&=\frac{i}{2m}d_6 \bar{\Psi} [D^\mu ,
f^+_{\mu\nu}]D^\nu \Psi+{\rm H.c.}+\frac{2 i}{m}d_7
\bar{\Psi} \left(\partial^\mu v^{(s)}_{\mu\nu}\right)D^\nu
\Psi+{\rm H.c.}+\cdots \:,
\end{split}
\end{equation}
where H.c.\ refers to the Hermitian conjugate and
\begin{align*}
f^\pm_{\mu\nu} &=u f_{L\mu\nu} u^\dagger \pm u^\dagger f_{R\mu\nu} u, \\
f_{R\mu \nu} &=\partial_\mu r_\nu- \partial_\nu r_\mu-i\left[r_\mu,r_\nu\right], \\
f_{L\mu \nu} &=\partial_\mu l_\nu- \partial_\nu
l_\mu-i\left[l_\mu,l_\nu\right], \\
v_{\mu\nu}^{(s)}&=\partial_\mu v_{\nu}^{(s)}-\partial_\nu
v_{\mu}^{(s)},
\end{align*}
with $r_\mu=l_\mu=-e  \mathcal{A}_\mu \tau_3/2$.

\subsection{Lagrangian containing vector mesons}

   The $\rho$-meson triplet consists of a pair of charged fields,
$\rho_\mu^\pm=(\rho_{1\mu}\mp i\rho_{2\mu})/\sqrt{2}$, and a third
neutral field, $\rho_\mu^0=\rho_{3\mu}$.
   Using a covariant Lagrangian formalism, self-interacting massive vector
fields are subject to constraints \cite{weinbergQTFI}.
   It was shown in Ref.~\cite{Djukanovic:2010tb} that the requirement for a
quantum field theory of vector mesons to be self consistent in terms of
constraints and perturbative renormalizability leads to relations among
the coupling constants of the Lagrangian.
   Eventually, at leading order the self-interacting part of the most
general effective Lagrangian for $\rho$ mesons reduces to a
massive Yang-Mills structure \cite{Djukanovic:2010tb},\footnote{
   From SU(2)-symmetry considerations alone, the interaction Lagrangian
would contain one three-vector and two four-vector interaction terms
with, in total, three independent coupling constants (see Eq.~(44)
of Ref.~\cite{Djukanovic:2010tb}).}
\begin{equation}
\label{Lagrho}
{\cal L}_{\rho\:\rm eff}=-\frac{1}{2}\textrm{Tr}\left(\rho_{\mu\nu}\rho^{\mu\nu}\right)
+M_\rho^2\textrm{Tr}(\rho_\mu\rho^\mu),
\end{equation}
where
\begin{align*}
\rho_\mu&=\rho_{k\mu}\frac{\tau_k}{2},\\
\rho_{\mu\nu}&=\partial_\mu \rho_\nu-\partial_\nu \rho_\mu-ig[\rho_\mu,\rho_\nu].
\end{align*}
   The Lagrangian of Eq.~(\ref{Lagrho}) contains two parameters, namely, the
$\rho$-meson mass $M_\rho$ (in the chiral limit) and a coupling strength $g$.
   Under the pair of local {\em chiral} transformations $(V_L,V_R)$, we choose the
$\rho$ mesons to transform inhomogeneously \cite{Weinberg:1968de}
(model III of Ref.\ \cite{Ecker:1989yg}),
\begin{equation}
\label{transformation_rho}
\rho_\mu\mapsto K\rho_\mu K^\dagger-\frac{i}{g}\partial_\mu K K^\dagger,
\end{equation}
where
\begin{displaymath}
K(V_L,V_R,U)={\sqrt{V_RUV_L^\dagger}}^{\,-1}V_R\sqrt{U}.
\end{displaymath}
   Equation (\ref{transformation_rho}) implies $\rho_{\mu\nu}\mapsto K \rho_{\mu\nu} K^\dagger$.
   The mass term remains chirally invariant through the replacement
$\rho_\mu\to\rho_\mu-(i/g)\Gamma_\mu$, the result of which transforms homogeneously
under local chiral transformations.
    Neglecting terms irrelevant for the calculation of the form factors,
the effective chiral Lagrangian can be written as
\begin{align}
\label{weinberg}
{\cal L}_{\pi\rho}={}&-\frac{1}{2}\textrm{Tr}\left(\rho_{\mu\nu}\rho^{\mu\nu}\right)
+M_\rho^2\textrm{Tr}\left[\left(\rho_{\mu}-\frac{i}{g}\Gamma_\mu\right)
\left(\rho^\mu-\frac{i}{g}\Gamma^\mu\right)\right]
+\frac{d_x}{2} \textrm{Tr}\left(\rho_{\mu\nu}f_+^{\mu\nu}\right)+\cdots.
\end{align}
   Besides the proper relations among the self couplings of the $\rho$ mesons, the Lagrangian of
Eq.~(\ref{weinberg}) gives rise to vector-meson dominance in the sense that, both, the $\rho\pi\pi$ and
the $\rho\gamma$ coupling contained in the mass term [second term on the right-hand side
of Eq.~(\ref{weinberg})] are of leading order.
   The $d_x$ term parameterizes a deviation of higher order.
   In model III of Ref.\ \cite{Ecker:1989yg} it is neglected, i.e., set to zero.
   According to Ref. \cite{Ecker:1989yg}, the Lagrangian of Eq.\ (\ref{weinberg}) is
obtained from the most general one by performing a field redefinition and implementing
certain relations among the coupling constants.
   These relations exactly correspond to those derived in Ref.\ \cite{Djukanovic:2010tb}
rendering Eq.\ (\ref{weinberg}) consistent with the constraints and renormalizable
in a perturbative sense.

   In addition to $\rho$ mesons, we also include the $\omega$ meson as
a dynamical degree of freedom.
   For our calculation, from the leading-order Lagrangian \cite{Ecker:1989yg}
we only need the coupling of the $\omega$ meson to external fields:
\begin{equation}
{\cal L}_{\pi\omega}^{(3)}=-f_{\omega} \left(\partial^\mu
\omega^\nu-\partial^\nu
\omega^\mu\right)v^{(s)}_{\mu\nu}+\cdots.\label{omega}
\end{equation}

   Finally, we require the coupling of vector mesons to the nucleon
which for our purposes is given by
\begin{align}
{\cal L}_{\pi V N}={}&\bar{\Psi}\left[g\left(\rho_\mu-\frac{i}{g}\Gamma_\mu\right)
+\frac{1}{2}\:g_{\omega}\:\omega_\mu\:\right]\gamma^\mu\:\Psi
+\frac{G_\rho}{2}\bar{\Psi}\rho_{\mu\nu}\sigma^{\mu\nu}
\Psi+\cdots.
\end{align}
    Here, we have applied the universality of the $\rho$-meson coupling $g_{\rho N N}=g$.
    In the realization of Ref.\ \cite{Weinberg:1968de},
the universal coupling is a consequence of chiral symmetry.
   In the present context, it is more likely to be a consequence of
consistency conditions imposed by the demand of perturbative
renormalizability \cite{Djukanovic:2004mm}.
   A coupling of the $\omega$ meson to the nucleon proportional to $\sigma^{\mu\nu}$ is
not needed at third chiral order, because there is no $\omega\gamma$
coupling at leading order in Eq.\ (\ref{omega}) as opposed to the
leading-order $\rho\gamma$ coupling of Eq.\ (\ref{weinberg}).

\subsection{Lagrangian containing the $\Delta(1232)$ resonance}

   The $\Delta(1232)$ resonance $\left[I(J^P)=\frac{3}{2}(\frac{3}{2}^+)\right]$ will be described by
a vector-spinor isovector-isospinor with components
\begin{eqnarray*}
 \Psi_{\mu,i}=
\left(
   \begin{array}{cc}
     \Psi_{\mu,i,\frac{1}{2}} \\
     \Psi_{\mu,i,-\frac{1}{2}}
   \end{array}
\right), \label{deltafeld}\qquad \mu=0,1,2,3, \qquad i=1,2,3.
\end{eqnarray*}
   The physical $\Delta$ consists of an isospin quadruplet, whereas the
description above involves six isospin components.
   In order to project onto the physical degrees of freedom, we introduce the isospin
projection operators (see, e.g., Sec.~4.7 of Ref.~\cite{Scherer:2012zzd} for more
details)
\begin{align*}
\xi^\frac{3}{2}_{ij,\alpha\beta}&=\delta_{ij}\delta_{\alpha\beta}
-\frac{1}{3}\left(\tau_i\tau_j\right)_{\alpha\beta},\\
\xi^\frac{1}{2}_{ij,\alpha\beta}&=\frac{1}{3}\left(\tau_i\tau_j\right)_{\alpha\beta},
\end{align*}
where the isovector components refer to a Cartesian isospin basis.
   Incorporating the projection operators explicitly, the leading-order
Lagrangian in $n$ space-time dimensions reads \cite{Hemmert:1997ye}
\begin{equation}
{\cal L}^{(1)}_{\pi\Delta}=\bar{\Psi}_\mu
\xi^\frac{3}{2}\Lambda^{(1)\mu\nu}_{\pi\Delta}(A,n)\xi^\frac{3}{2}\Psi_{\nu}\label{delta},
\end{equation}
with
\begin{align}
\Lambda^{(1)\mu\nu}_{\pi\Delta}(A,n)={}&-\Big\{(i\slashed{D}-m_\Delta)g^{\mu\nu}
+iA(\gamma^\mu D^\nu+\gamma^\nu D^\mu)\nonumber\\
&+\frac{i}{n-2}\left[(n-1)A^2+2A+1\right]\gamma^\mu\slashed{D}\gamma^\nu\nonumber\\
&+\frac{m_\Delta}{(n-2)^2}\left[n(n-1)A^2+4(n-1)A+n\right]\gamma^\mu\gamma^\nu\nonumber\\
&+\frac{\texttt{g}_1}{2}\slashed{u}\gamma_5g^{\mu\nu}+\frac{\texttt{g}_2}{2}(\gamma^\mu
u^\nu+u^\mu\gamma^\nu)\gamma_5+\frac{\texttt{g}_3}{2}\gamma^\mu\slashed{u}\gamma_5\gamma^\nu\Big\}.
\label{LambdapiDelta}
\end{align}
   Here, the covariant derivative is given by
\begin{align*}
\left(D_\mu \Psi\right)_{\nu,i,\alpha} &= {\cal
D}_{\mu,ij,\alpha\beta}\Psi_{\nu,j,\beta}, \\ \nonumber {\cal
D}_{\mu,ij,\alpha\beta}
 &=\partial_\mu \delta_{ij}\delta_{\alpha \beta}-2i\epsilon_{ijk}
 \Gamma_{\mu,k}\delta_{\alpha\beta}
 +\delta_{ij}\Gamma_{\mu,\alpha\beta}-iv_\mu^{(s)}\delta_{ij}\delta_{\alpha
 \beta},
\end{align*}
where we parameterized $\Gamma_\mu=\Gamma_{\mu,k}\tau_k$.
   In Eq.\ (\ref{delta}), $A\neq-\frac{1}{2}$ denotes an arbitrary real
parameter and $m_\Delta$ refers to the leading-order mass of the
$\Delta$.

   Since the Lagrangian of Eq.\ (\ref{delta}) describes a system with
constraints, similarly to the previously discussed case of vector
mesons, the requirement for a self-consistent theory leads to
relations among the coupling constants \cite{Wies:2006rv},
\begin{equation*}
\texttt{g}_2=A\texttt{g}_1, \qquad \texttt{g}_3=-\frac{1+2A+A^2(n-1)}{n-2}\:\texttt{g}_1.
\end{equation*}

   The lowest-order $\pi N \Delta$ interaction Lagrangian reads
\cite{Hacker:2005fh},
\begin{equation}
{\cal L}^{(1)}_{\pi N\Delta}
=\texttt{g}\bar{\Psi}_{\mu,i}\xi_{ij}^\frac{3}{2}\left(g^{\mu\nu}
+\frac{1+3A}{2}\:\gamma^\mu\gamma^\nu\right)u_{\nu,j}\Psi+{\rm H.c.}\:,
\label{LpiNDelta}
\end{equation}
with the parameterization $u_\mu= u_{\mu,k}\tau_k$, and $\texttt{g}$
being a coupling constant.\footnote{The sign convention in
Eq.~(\ref{LpiNDelta}) is chosen such that SU(4) symmetry implies the
relations $\texttt{g}_1=\frac{9}{5}\texttt{g}_A$ and
$\texttt{g}=\frac{3}{5}\sqrt{2}\texttt{g}_A$ among the coupling
constants of Eqs.~(\ref{LpiN}), (\ref{LambdapiDelta}), and
(\ref{LpiNDelta}) \cite{Hemmert:1997ye}.}
   Since physical quantities cannot depend on $A$ \cite{Hacker:2005fh},
we choose $A=-1$ in the following calculations.

\subsection{Power counting}
\label{power_counting}
   We assign a low-energy order $D$ to each renormalized diagram.
   The value of $D$ is determined with the following power-counting rules:
   A pion propagator counts as $\mathcal{O}(q^{-2})$, a nucleon propagator
as $\mathcal{O}(q^{-1})$, vertices derived from ${\cal L}_{2}$ count as
$\mathcal{O}(q^{2})$, and vertices from ${\cal L}_{\pi N}^{(i)}$
count as $\mathcal{O}(q^{i})$.
   Both $\rho$-meson and $\omega$-meson propagators count as $\mathcal{O}(q^{0})$
while vertices from ${\cal L}_{\pi\omega}^{(3)}$ count as $\mathcal{O}(q^{3})$.
   From the listed terms of ${\cal L}_{\pi\rho}$ and ${\cal L}_{\pi VN}$ vertices of
$\mathcal{O}(q^{0})$ to $\mathcal{O}(q^{3})$ can be derived.
   The $\Delta$ propagator counts as $\mathcal{O}(q^{-1})$ and vertices
from ${\cal L}^{(1)}_{\pi N\Delta}$ count as $\mathcal{O}(q)$.
   Finally, we assign the order $\mathcal{O}(q)$ to the mass difference
$\delta\equiv m_\Delta-m$.
   In order to renormalize the loop diagrams in such a way that they respect
the above power counting, we apply the EOMS scheme
\cite{Fuchs:2003qc}.

\section{Results and Discussion}

\begin{figure}
\begin{center}
\includegraphics[width=0.65\textwidth]{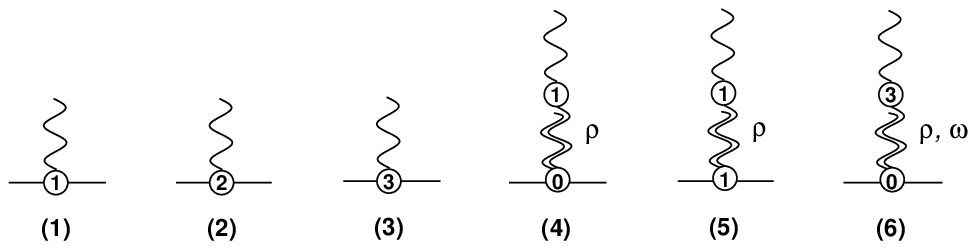}\\
\vspace{0.1 cm}
\includegraphics[width=0.8\textwidth]{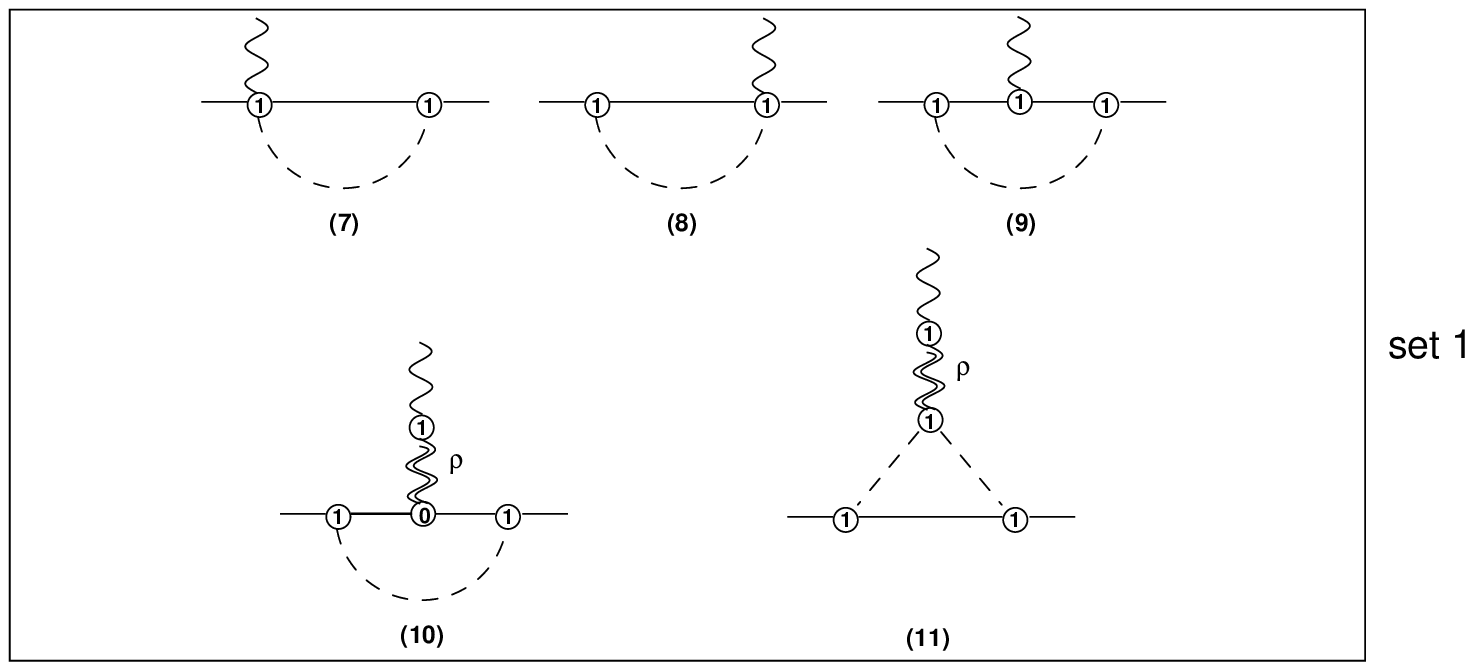}\\
\vspace{0.1 cm}
\includegraphics[width=0.8\textwidth]{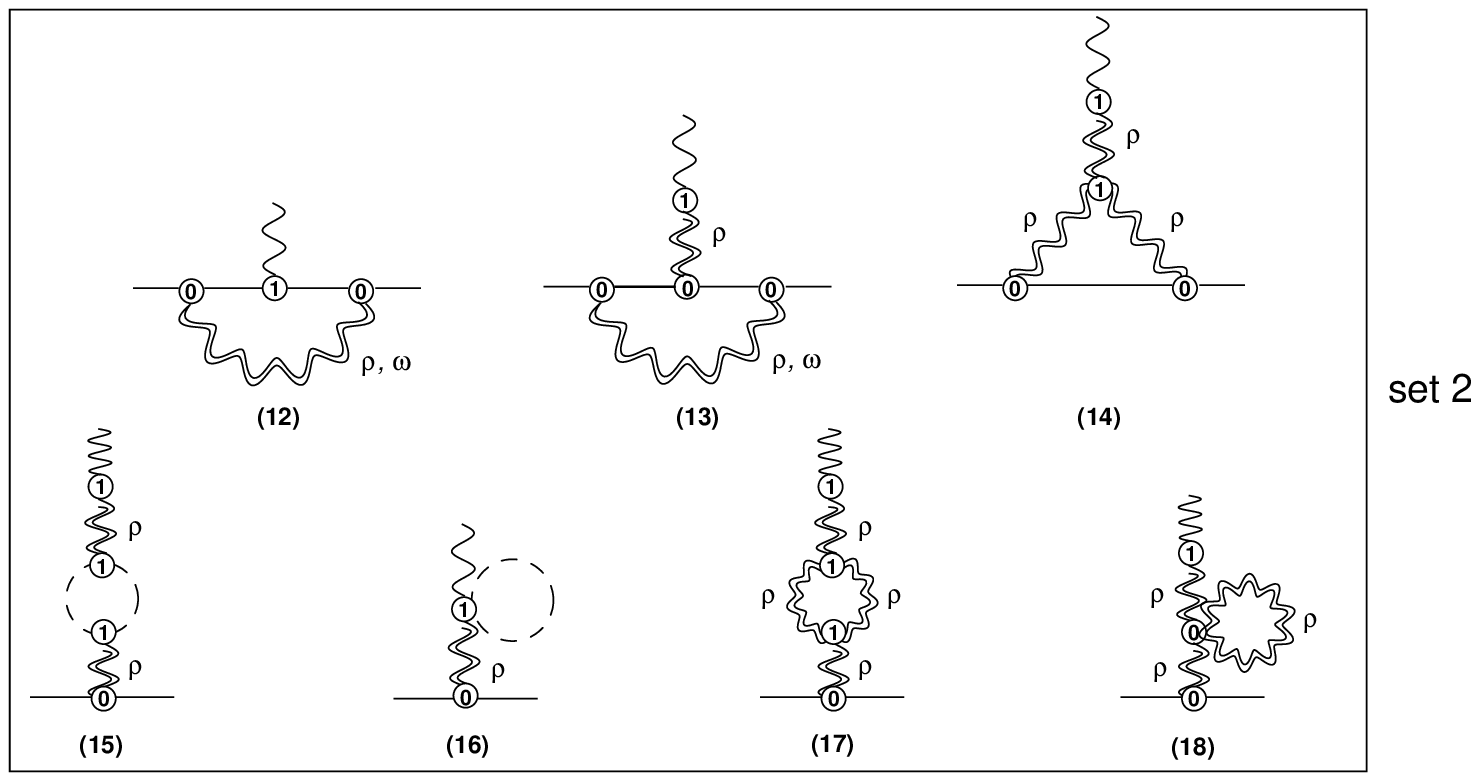}\\
\vspace{0.1 cm}
\includegraphics[width=0.8\textwidth]{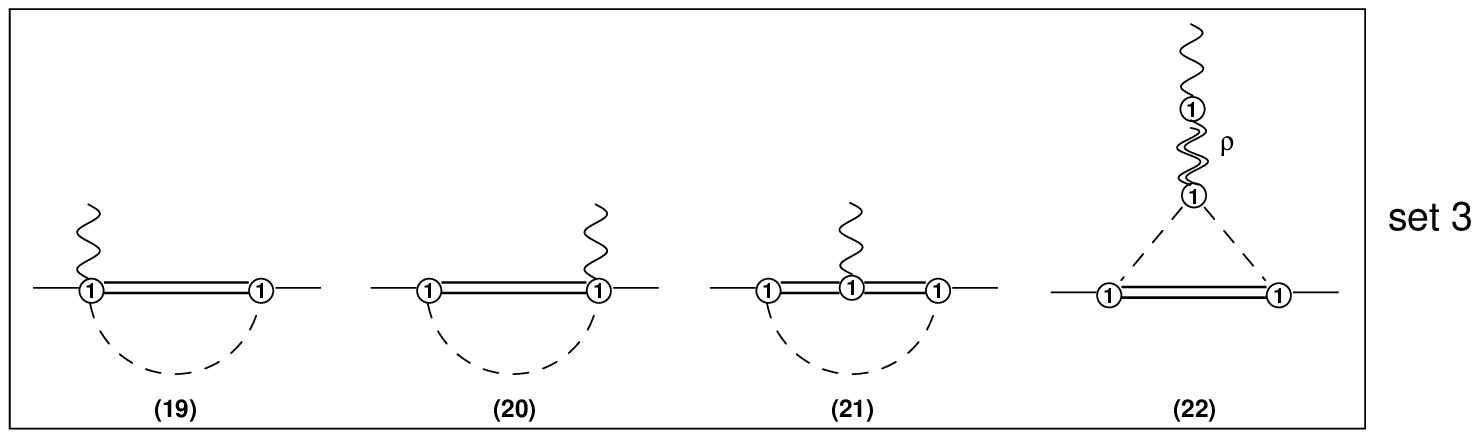}
\end{center}
\caption{Feynman diagrams including vector mesons and the $\Delta$
resonance contributing to the electromagnetic form factors of
the nucleon up to and including $\mathcal{O}(q^3)$. Solid, dashed,
wiggly, double-wiggly, and double-solid lines refer to nucleons,
photons, pions, vector mesons, and the $\Delta$ resonance, respectively. The
numbers in the interaction blobs denote the chiral order of the
respective vertex.}\label{feynman}
\end{figure}

   All Feynman graphs contributing to the calculation of the
electromagnetic form factors up to and including
$\mathcal{O}(q^{3})$ are displayed in Fig.~\ref{feynman}.\footnote{After renormalization,
the diagrams (14), (17), and (18) are at least of $\mathcal{O}(q^{5})$ and can thus be
neglected in the numerical analysis.}
   The 16 loop diagrams of Fig.~\ref{feynman} are grouped into three, independently
current-conserving subsets.
   In the following, we refer to the diagrams (7)--(11) as set 1, the diagrams (12)--(18)
as set 2, and the diagrams (19)--(22) as set 3, respectively.
   Set 1 consists of diagrams proportional to $\texttt{g}_A^2/F^2$ containing only pion loops
while set 2 consists of diagrams proportional to $g^2$ and
$g_\omega^2$ containing pion loops as well as vector-meson loops.
   Finally, set 3 contains all pion-loop diagrams involving the $\Delta$
resonance, thus being proportional to $\texttt{g}^2$.

   Summing up all contributions and multiplying them with the wave
function renormalization constant $Z_N$ yields the final expressions
for the form factors. To render the results for the unrenormalized
form factors finite, we apply the modified minimal subtraction
scheme of ChPT ($\widetilde{\rm MS}$) \cite{Gasser:1983yg}.
   Beyond that, we perform finite subtractions according to the EOMS scheme
\cite{Fuchs:2003qc} such that the power counting of Sec.\ \ref{power_counting} is respected.
   To the given order, the product of the wave function renormalization constant
and the tree-order diagrams subtracts all power-counting-violating terms of the loop
diagrams in the Dirac form factor $F_1$.
   In agreement with the Ward identity, we obtain $F_1^p(0)=1$
and $F_1^n(0)=0$ for the proton and neutron, respectively.
   On the other hand, the loop contributions to the Pauli form factor $F_2$
contain power-counting-violating terms.
   All subtraction terms are analytic in the pion mass and momenta and can be
absorbed in the renormalization of the available coupling constants.

\subsection{Fixing of the LECs}

   To evaluate the form factors numerically, the
parameters of the effective Lagrangian need to be fixed.
   The masses, the axial-vector coupling constant, and the pion-decay
constant are expressed in terms of their physical values, because the
difference to the respective values in the chiral limit is beyond the
accuracy of our calculation.

   Using the KSRF relation \cite{Kawarabayashi:1966kd,Riazuddin:1966sw},
\begin{equation}
M_\rho^2=2g^2F^2,
\label{ksrf}
\end{equation}
generated by the combination of chiral symmetry and
the consistency of the EFT with respect to renormalizability
\cite{Djukanovic:2004mm}, we obtain $g=5.93$.
   Moreover, we take $\texttt{g}=1.13$ as obtained from a fit to the $\Delta \rightarrow
\pi N$ decay width \cite{Hacker:2005fh}.
   The numerical values of the above parameters are summarized in Table \ref{masses}.

\begin{table}
\begin{center}
\begin{tabular}{|c|c|c|c|c|c|c|c|c|}
\hline
$m_N$ & $M_\pi$ & $M_\rho$ & $M_\omega$ & $m_\Delta$ & $F_\pi$ &  $\texttt{g}_A$&  $g$ &   $\texttt{g}$  \\
\hline\hline
    \quad0.938\quad\quad & \quad0.140\quad\quad & \quad0.775\quad\quad  & \quad0.783\quad\quad &
    \quad1.21\quad\quad & \quad0.0922\quad\quad & \quad1.27 \quad\quad & \quad5.93 \quad\quad& \quad1.13 \quad\quad   \\
\hline
\end{tabular}
\end{center}
\caption{Input parameters: The masses and $F_\pi$ are given in units
of GeV; the coupling constants $\texttt{g}_A$, $g$, and $\texttt{g}$
are dimensionless. \label{masses}}
\end{table}

   To determine the renormalized low-energy constants $c_6$ and $c_7$,
we fix the Pauli form factors $F_2(Q^2)$ at $Q^2=0$ in accordance
with Eq.\ (\ref{magneticmoments}).
   The expansions of the anomalous magnetic moments of the proton and neutron read
\begin{equation}
\label{kappa}
\begin{split}
\kappa_p &= c_7 m + 2 m \left( c_6  - \frac{G_\rho}{2g}\right) - \frac{\texttt{g}_A^2m}{8 F^2 \pi}M
+\frac{\texttt{g}^2 m }{9 F^2 \pi^2}\:\delta\left[\ln\left(\frac{M}{2\delta}\right)
-\frac{\sqrt{\delta^2-M^2}}{\delta}\ln\left(X\right)\right]+\cdots,\\
\kappa_n &= c_7 m - 2 m \left( c_6  - \frac{G_\rho}{2g}\right) + \frac{\texttt{g}_A^2 m }{8 F^2 \pi}M
-\frac{\texttt{g}^2 m }{9 F^2
\pi^2}\:\delta\left[\ln\left(\frac{M}{2\delta}\right)
-\frac{\sqrt{\delta^2-M^2}}{\delta}\ln\left(X\right)\right]+\cdots,
\end{split}
\end{equation}
with
\begin{equation}
\label{definition_X} X=\frac{\delta-\sqrt{\delta^2-M^2}}{M}\;,\qquad
 \delta=m_\Delta-m.
\end{equation}
   In Eq.~(\ref{kappa}), the ellipses refer to terms scaling at least as $t^2$ under
$M\mapsto t M$ and $\delta\mapsto t \delta$.
   Because of the chosen renormalization scheme, the Feynman diagrams
of set 2 do not contribute to the magnetic moments.
   The non-analytic terms for the magnetic moments of Eq.\ (\ref{kappa}) coincide
with those of Ref. \cite{Bernard:1998gv}.
   Using the values of Table \ref{masses} for the input parameters, the
$\pi N$ and $\pi\Delta$ loop contributions to the isoscalar and isovector magnetic moments
are shown in Table \ref{table_magnetic_moments}.
   Keeping only the leading-order terms, the pion loop contributions are purely
isovector as in ordinary ChPT at ${\cal O}(q^3)$.
   On the other hand, evaluating the full expressions modifies the
isovector pieces and also generates isoscalar contributions.
\renewcommand{\arraystretch}{1.3}
\begin{table}
\begin{center}
\begin{tabular}{|l||c|c|c|c|}
\hline
& $\kappa_{\pi N}^{(s)}$ & $\kappa_{\pi N}^{(v)}$ & $\kappa_{\pi \Delta}^{(s)}$ & $\kappa_{\pi \Delta}^{(v)}$\\
\hline\hline
Expanded & \quad 0 \quad\quad & \quad 1.98 \quad\quad & \quad 0 \quad\quad & \quad $-0.222$ \quad\quad\\
\hline
Full & \quad 0.169 \quad\quad & \quad 1.35 \quad\quad & \quad $-0.0120$ \quad\quad  & \quad $-0.150$ \quad\quad \\
\hline
\end{tabular}
\end{center}
\caption{$\pi N$ and $\pi\Delta$ loop contributions to the isoscalar and isovector anomalous magnetic moments.}
\label{table_magnetic_moments}
\end{table}
\renewcommand{\arraystretch}{1}

   Adjusting the complete results for the magnetic moments to their empiric values yields
\begin{equation}
\tilde{c}_6=1.39\,\textrm{GeV}^{-1} , \qquad c_7=-0.148\:
\textrm{GeV}^{-1},
\end{equation}
with $\tilde{c}_6\equiv c_6  - G_\rho/(2g)$.
   In Sec.~\ref{graphical}, we compare our results with explicit
$\Delta$ contributions to those without the $\Delta$ resonance.
   Therefore, we also state the values for the couplings of the latter case, namely,
\begin{equation}
\tilde{c}_6=1.35\, \textrm{GeV}^{-1}, \qquad c_7=-0.154\:
\textrm{GeV}^{-1}.
\end{equation}

   The remaining six free low-energy coupling constants $G_\rho$,
$f_\omega$, $g_\omega$, $d_6$, $d_7$, and $d_x$ are determined by
simultaneous fits of all four Sachs form factors to experimental
data for different regions of momentum transfer. As the data basis
for the fits, we use the extensive proton cross section data set
from Refs.\ \cite{Bernauer:2010wm,bernauerphd} and the neutron form
factor data from Refs.\
\cite{Hanson:1973vf,JonesWoodward:1991ih,Thompson:1992ci,Markowitz:1993hx,Gao:1994ud,Anklin:1994ae,Eden:1994ji,
Bruins:1995ns,Anklin:1998ae,Ostrick:1999xa,Passchier:1999cj,Herberg:1999ud,Becker:1999tw,Xu:2000xw,Kubon:2001rj,Xu:2002xc,Glazier:2004ny,Anderson:2006jp,
Geis:2008aa}.
   Because of the small numerical contributions originating from the diagrams
of set 2, the fits depend only marginally on the individual values
of $g_\omega$ and $f_\omega$.
   On the other hand, the product of $g_\omega$ and $f_\omega$
stemming from the tree diagram (6) of Fig.\ \ref{feynman} is much
much more influential for the final result.
   Thus we fix $f_\omega$ at 0.1 and only use the product
$f_\omega\cdot g_\omega$ as an independent fit parameter.
   The results for the fitted renormalized couplings are shown in
Table \ref{fit}.

\begin{table}
\begin{center}
\begin{tabular}{|c||c|ccccc c|c|}
\hline Resonances&$Q^2_{\textrm{max}}$& $\quad g_\omega$ \quad&
$\quad f_\omega\cdot g_\omega$ \quad&\quad $G_\rho$ \quad&\quad $d_6$ \quad&\quad $d_7$ \quad&\quad$d_x$\quad&\quad$\chi_\textrm{red}^2$\quad \\
\hline\hline
\multirow{4}{*}{$\Delta$, $\rho$, $\omega$ }&0.2 &\quad $-1.06$\quad&\quad $-0.106$\quad&\quad $-4.84$\quad&\quad 1.67&\quad $-0.282$\quad&\quad$-0.506$\quad&\quad1.50\quad\\
&0.3 &\quad $-1.27$ \quad&\quad $-0.127$\quad&\quad $-4.05$\quad&\quad 1.55\quad&\quad $-0.233$\quad&\quad$-0.512$ \quad&\quad4.21\quad\\
&0.4 &\quad $-1.92$ \quad&\quad $-0.192$\quad&\quad $-1.98$\quad&\quad 1.50\quad&\quad $-0.211$\quad&\quad$-0.547$ \quad&\quad25.30\quad\\
\hline
\multirow{4}{*}{$\rho$, $\omega$}&0.2 &\quad 5.13\quad&\quad 0.513\quad&\quad $-16.90$\quad&\quad 0.629\quad&\quad 0.0909\quad&\quad$-0.134$\quad&\quad1.45\quad \\
&0.3 &\quad 4.91\quad&\quad 0.491\quad&\quad $-17.13$\quad&\quad 0.507\quad&\quad 0.0991\quad&\quad$-0.118$ \quad&\quad1.74\quad \\
&0.4 &\quad 4.49\quad&\quad 0.449\quad&\quad $-16.58$\quad&\quad 0.490\quad&\quad 0.0934\quad&\quad$-0.130$ \quad&\quad3.57\quad \\
    \hline
\end{tabular}
\end{center}
\caption{Comparison of the renormalized coupling constants obtained
from fits of the results to different ranges of momentum transfer.
    The second set includes the $\Delta$ resonance as an explicit dynamical degree of freedom.
$G_\rho$ is
given in units of GeV$^{-1}$, $d_6$ and $d_7$ in units of
GeV$^{-2}$, and $Q^2_{\textrm{max}}$ in units of GeV$^{2}$; the
remaining coupling constants $d_x$, $g_\omega$, and $f_\omega$ are
dimensionless.\label{fit}}
\end{table}

   As indicated by the respective values of the reduced chi-square test
$(\chi_\textrm{red}^2\equiv\chi^2/\textrm{d.o.f.})$, the adjusted
results including only vector mesons as explicit resonant degrees of
freedom show better agreement with experimental data than the
results incorporating also the $\Delta$ resonance.
   As the range of momentum transfer increases, the respective values for
$\chi_\textrm{red}^2$ of the fits including the $\Delta$ increase
faster than those without the $\Delta$.
   We will come back to this feature in Secs.~\ref{graphical} and ~\ref{estimate}.

\subsection{Charge and magnetic radii}

   Expanding the Dirac and Pauli form factors for small values of $Q^2$,
\begin{displaymath}
F_i^{(s,v)}(Q^2)=F_i^{(s,v)}(0)\left(1-\frac{1}{6} \langle (r_i^{(s,v)})^2\rangle Q^2 + \cdots\right),
\end{displaymath}
gives access to the mean square radii.
   At ${\cal O}(q^3)$, the expanded mean square radii are given by
\begin{align}
\label{r1s2}
\langle(r_1^{(s)})^2\rangle&=-24 d_7 +\frac{12 c_\omega}{M_\omega^2}
+\frac{9 \texttt{g}_A^2}{32 F^2\pi^2}
+\frac{\texttt{g}^2}{288 F^2\pi^2}\left[-17+40\ln\left(\frac{m}{\mu}\right)\right]\nonumber\\
&\quad-3g^2f\left(M_\rho\right)-g_\omega^2 f\left(M_\omega\right)+\cdots,\\
\label{r1v2}
\langle (r_1^{(v)})^2\rangle&=-12 d_6+6\frac{1-d_xg}{M_\rho^2}\nonumber\\
&\quad-\frac{1}{16 F^2 \pi ^2}\left[2 \ln\left(\frac{M}{\mu}\right)+1\right]
-\frac{\texttt{g}_A^2}{16 F^2 \pi^2}\left[10 \ln\left(\frac{M}{\mu}\right)
-12 \ln \left(\frac{m}{\mu}\right)+\frac{41}{2}\right]\nonumber\\
&\quad+\frac{\texttt{g}^2}{54 F^2 \pi ^2}
   \left\{\frac{379}{16}-10 \ln \left(\frac{m}{\mu}\right)
   -\frac{3m^2}{16 M_\rho^2} \left[60 \ln \left(\frac{m}{\mu}\right)+7\right]
   +30 \ln \left(\frac{M}{\mu}\right)
   \right.\nonumber\\
   &\quad-\left.\frac{30 \delta  \ln \left(X\right)}{\sqrt{\delta
   ^2-M^2}}\right\}+g^2f\left(M_\rho\right)-g_\omega^2f\left(M_\omega\right)+\cdots,\\
\langle(r_2^{(s)})^2\rangle&=0+\cdots,\\
\langle(r_2^{(v)})^2\rangle&=\frac{\texttt{g}_A^2 m}{8 F^2 \pi M
\kappa^{(v)}}-\frac{\texttt{g}^2 m}{9 F^2 \pi^2
\sqrt{\delta^2-M^2}\kappa^{(v)}}\ln\left(X\right)+\cdots,
\end{align}
where $X$ is defined in Eq.\ (\ref{definition_X}).
   In the case of the mean square Dirac radii, the ellipses refer to terms that scale at least
linearly in $t$ under $M\mapsto t M$ and $\delta\mapsto t \delta$.
   On the other hand, for the mean square Pauli radii, the ellipses represent
terms that remain constant or scale with higher powers.
   As expected \cite{Beg:1973sc}, $\langle (r_1^{(v)})^2\rangle$ diverges logarithmically
in the chiral limit, whereas $\langle (r_2^{(v)})^2\rangle$ shows a $1/M$ singularity.
  The respective contributions of the vector mesons are given in terms of the
function
\begin{align}
f\left(M_V\right)={}&\frac{12 m^4-37 M_V^2 m^2+10 M_V^4}{64 m^4
\left(4 m^2-M_V^2\right)\pi^2}+\frac{\left(4 m^4-6 M_V^2 m^2+5
M_V^4\right) \ln
\left(\frac{M_V}{m}\right)}{32 m^6 \pi^2}\nonumber\\
&+\frac{M_V \left(-36 m^6+70 M_V^2 m^4-36 M_V^4 m^2+5 M_V^6\right)
\arccos \left(\frac{M_V}{2 m}\right)}{32
   m^6 \left(4 m^2-M_V^2\right)^{3/2}\pi^2},
\label{fVM}
\end{align}
which vanishes in the limit of infinitely heavy vector-meson masses,
\begin{align}
\label{limesfVM} \lim_{M_V\to\infty} f\left(M_V\right)=0.
\end{align}

   Using the couplings from the fitting procedure (see Table \ref{fit}),
we are in the position to determine the numerical values for the mean square charge and
magnetic radii, defined as
\begin{align}
\nonumber\langle\left(r_E^p\right)^2\rangle&=\left.-\frac{6}{ G_E^p
(0)}\frac{\textrm{d} G_E^p (Q^2)}{\textrm{d}Q^2}\right|_{Q^2=0},&
\langle\left(r_E^n\right)^2\rangle&=\left.-6\frac{\textrm{d}
G_E^n (Q^2)}{\textrm{d}Q^2}\right|_{Q^2=0},\\
\langle\left(r_M^p\right)^2\rangle&=\left.-\frac{6}{ G_M^p
(0)}\frac{\textrm{d} G_M^p (Q^2)}{\textrm{d}Q^2}\right|_{Q^2=0},&
\langle\left(r_M^n\right)^2\rangle&=\left.-\frac{6}{ G_M^n
(0)}\frac{\textrm{d} G_M^n (Q^2)}{\textrm{d}Q^2}\right|_{Q^2=0}.
\label{defRadii}
\end{align}
   The respective empirical values are shown together with our results in Table
\ref{radien}.\footnote{When evaluating numerical expressions, we
make use of the renormalization scale $\mu=1$ GeV.}
   As a general trend we find that the proton radii are better described
in terms of the calculation including the $\Delta$ resonance.
   In contrast, the neutron radii are in better agreement with the experimental
results in the theory without the $\Delta$ resonance.
   In all cases, $\langle\left(r_M^n\right)^2\rangle$ is smaller and
$|\langle\left(r_E^n\right)^2\rangle|$ is larger than their
respective empirical values.
   The situation improves in the calculation featuring just vector mesons as
$\langle\left(r_M^n\right)^2\rangle$ is approximately in agreement with experiment.
   Again, the discrepancy especially for
$\langle\left(r_M^n\right)^2\rangle$ grows towards larger
$Q^2_{\textrm{max}}$.
   Even though the value of $|\langle\left(r_E^n\right)^2\rangle|$ is notably smaller than in the
results with an explicit $\Delta$ resonance, it is still larger than
empirically predicted.

   In principle, one could adjust $d_6$ and $d_7$ to the electric radii
and two LECs originating from the vector-meson Lagrangian to the magnetic
radii, respectively.
   However, such an approach would be against the purpose/spirit of introducing
vector mesons, namely, generating curvature to extend the description to
intermediate values of $Q^2$.

   Finally, in Table \ref{table_mean_square_radii} we display the individual
$\pi N$, $\pi\Delta$, and vector-meson loop contributions to the mean square radii.
   The parameters have been taken from Table \ref{masses}.
   Note that the $\omega$-meson loop contributes only to the mean square radii of
the proton.
   The contribution is given by the second term in the respective sum and
has been evaluated with the largest value of Table \ref{fit},
i.e.~$g_\omega=5.13$.
   The difference between the full and expanded loop results may be regarded as
an, admittedly, rough error estimate.

\renewcommand{\arraystretch}{1.3}
\begin{table}
\begin{center}
\begin{tabular}{|c||c|c c c c|}
\hline Resonances&$Q^2_{\textrm{max}}$& $\quad
\langle\left(r_E^p\right)^2\rangle$ \quad&
\quad $\langle\left(r_E^n\right)^2\rangle$ \quad&\quad $\langle\left(r_M^p\right)^2\rangle$ \quad&\quad $\langle\left(r_M^n\right)^2\rangle$ \quad\\
\hline\hline
\multirow{4}{*}{$\Delta$, $\rho$, $\omega$ }
&$0.2$ &\quad $0.740$\quad&\quad $-0.288$\quad&\quad $0.631$\quad&\quad $0.718$\quad\\
&$0.3$ &\quad $0.744$ \quad&\quad $-0.355$\quad&\quad $0.614$\quad&\quad $0.700$\quad\\
&$0.4$ &\quad $0.761$ \quad&\quad $-0.440$\quad&\quad $0.574$\quad&\quad $0.667$\quad\\
\hline \multirow{4}{*}{$\rho$, $\omega$}
&$0.2$ &\quad $0.733$\quad&\quad $-0.198$\quad&\quad $0.677$\quad&\quad $0.725$\quad\\
&$0.3$&\quad $0.730$\quad&\quad $-0.221$\quad&\quad $0.672$\quad&\quad $0.728$\quad\\
&$0.4$ &\quad $0.734$\quad&\quad $-0.252$\quad&\quad $0.659$\quad&\quad $0.726$\quad\\
    \hline\hline
\multicolumn{2}{|c|}{Empirical values}&\quad $0.770$\quad&\quad $-0.116$\quad&\quad $0.604$\quad&\quad $0.743$\quad\\
 \hline
\end{tabular}
\end{center}
\caption{Comparison of the mean square charge and magnetic radii of the nucleon
obtained from the form factor results including and excluding the
$\Delta$ resonance fitted to different ranges of momentum transfer.
The mean square radii are given in units of fm$^{2}$. The empirical values are
taken from Ref. \cite{Beringer:1900zz}. \label{radien}}
\end{table}
\renewcommand{\arraystretch}{1}

\renewcommand{\arraystretch}{1.3}
\begin{table}
\begin{center}
\begin{tabular}{|l||cccc|}
\hline
Mean square radii &$\langle\left(r_E^p\right)^2\rangle$&$\langle\left(r_E^n\right)^2\rangle$&
$\langle\left(r_M^p\right)^2\rangle$&$\langle\left(r_M^n\right)^2\rangle$\\
\hline
\hline
$\pi N$, expanded & 0.304 & $-0.0938$ & 0.301 & 0.381\\
$\pi N$, full & 0.365 & $-0.158$ & 0.162 & 0.201\\
\hline
$\pi\Delta$, expanded & 0.0312 & $-0.0713$ & 0.0796 & 0.0517\\
$\pi\Delta$, full & $-0.115$ & 0.163 & 0.154 & 0.0323 \\
\hline\hline
Vector-meson loops, expanded & $0.0111$ & 0.00554  & $-0.000953$ & 0.00278 \\
&$+0.00821$&&$+0.00311$&\\
Vector-meson loops, full & $0.0422$ & $-0.00897$  & $-0.0153$ & 0.00400\\
&$+0.00821$&&$+0.00311$&\\
\hline
\end{tabular}
\end{center}
\caption{$\pi N$, $\pi\Delta$, and vector-meson loop contributions to the mean square radii
in units of fm$^{2}$.
The omega loop contributes only to the proton radii and is given by the second term in
the sum.
}
\label{table_mean_square_radii}
\end{table}
\renewcommand{\arraystretch}{1}

\subsection{Graphical representation of the form factor results}
\label{graphical}

\begin{figure}[t]
\begin{center}
\includegraphics[width=\textwidth]{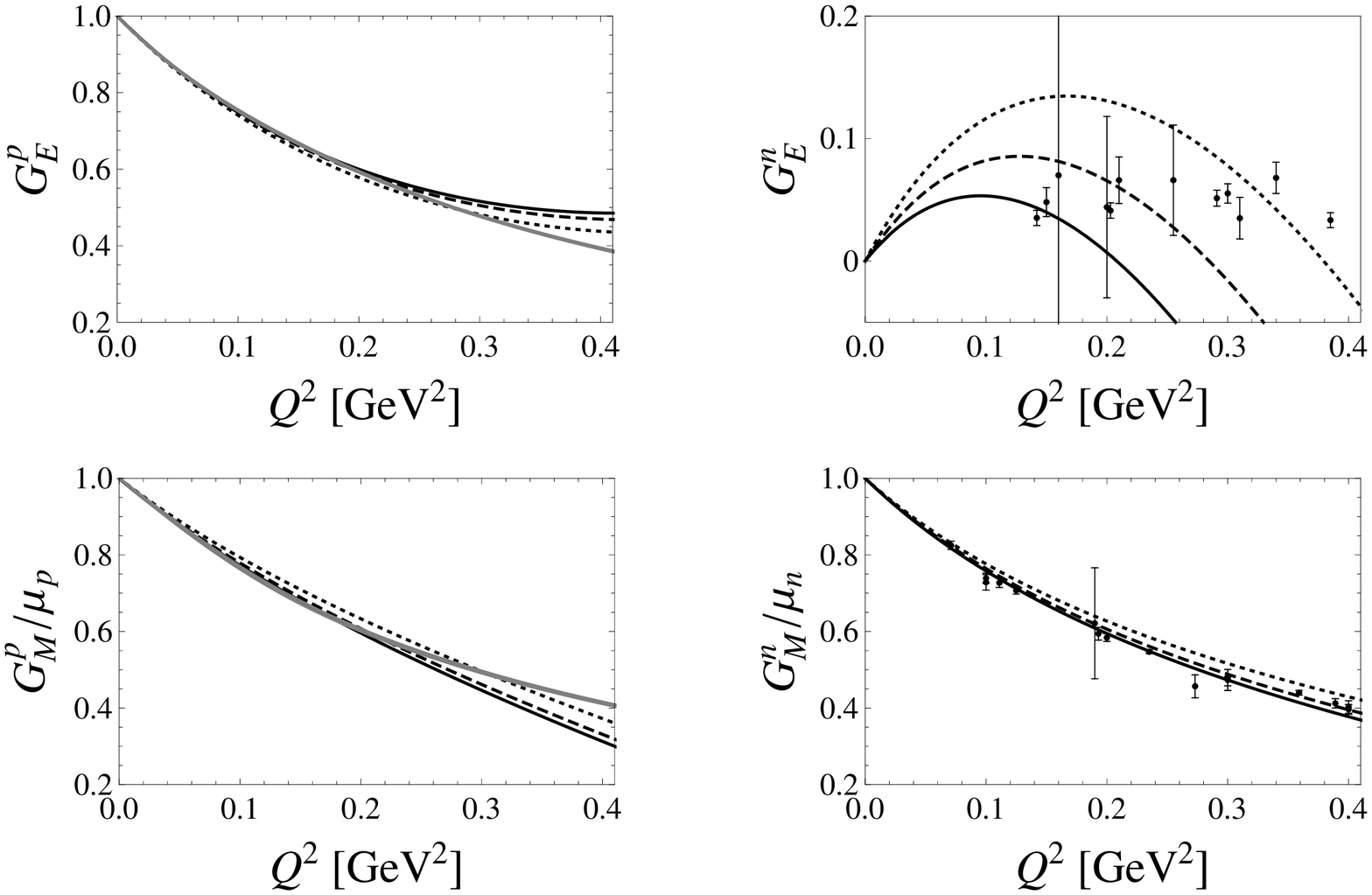}
\end{center}
\caption{Sachs form factors of the nucleon at $\mathcal{O}(q^3)$
including $\rho$, $\omega$, and $\Delta$, fitted to different ranges
of momentum transfer $Q^2$. Full lines correspond to a fit up to
$Q^2_\textrm{max}=0.2$ $\textrm{GeV}^2$, dashed lines up to
$Q^2_\textrm{max}=0.3$ $\textrm{GeV}^2$, and dotted lines up to
$Q^2_\textrm{max}=0.4$ $\textrm{GeV}^2$, respectively. The grey
bands represent empirical fits of the form factors to the measured
cross sections. } \label{DeltaFFQvar}
\end{figure}

   Using the LECs represented in Tables \ref{masses} and \ref{fit}, the
final results for the Sachs form factors are displayed in Fig.
\ref{DeltaFFQvar}. Since our proton results are fitted directly to
cross sections we show a grey band corresponding to a direct
least-squares model fit for $G_E^p$ and $G_M^p$ to the measured
cross sections and thus representing the experimental data
\cite{Bernauer:2010wm}. For the neutron form factors our results are
plotted together with the respective set of data to which they have
been fitted to. All curves describe the corresponding experimental
data reasonably well for all four form factors.
   As indicated by the values of $\chi_\textrm{red}^2$ in Table \ref{fit},
the description of the form factors adjusted to a wider range of momentum
transfer worsens.
   While $G^p_E$ and $G^n_M$ are still in good agreement with the corresponding
data for $Q^2_\textrm{max}=0.4$ $\textrm{GeV}^2$ (dotted curves),
$G^p_M$ does not show sufficient curvature.
   The electric form factor of the neutron, $G^n_E$, cannot be described well
beyond $Q^2\approx0.3$ $\textrm{GeV}^2$.
   In all cases, the slope of $G^n_E$ at small values of the momentum transfer
turns out too big.

\begin{figure}[t]
\begin{center}
\includegraphics[width=\textwidth]{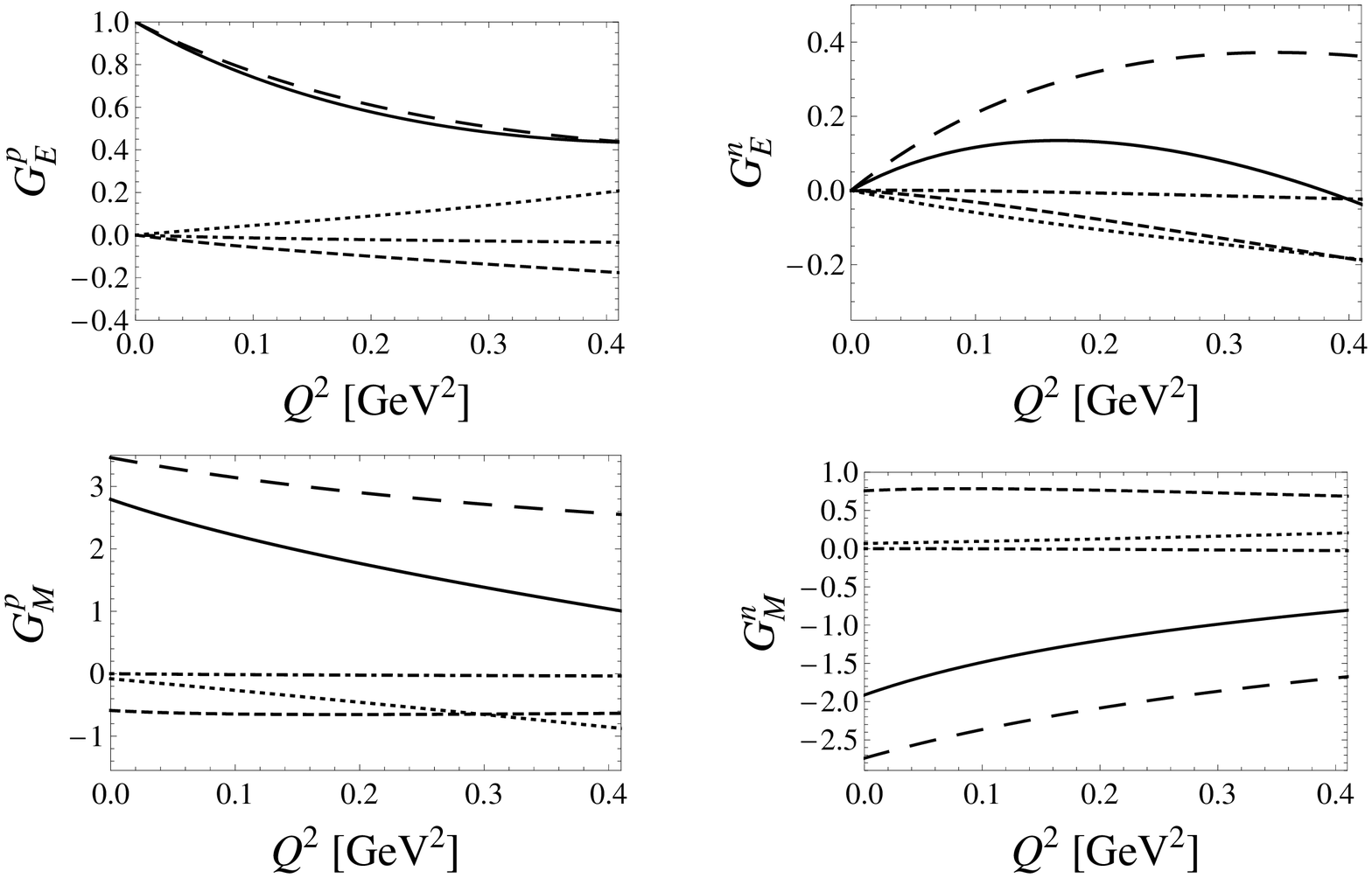}
\end{center}
\caption{Decomposition of the Sachs form factors of the nucleon at
$\mathcal{O}(q^3)$ including vector mesons and the $\Delta$
resonance. Full lines: total results; long-dashed lines: tree
contribution; short-dashed lines: contribution of set 1;
dash-dotted lines: contribution of set 2; dotted lines: contribution
of set 3 of Fig.\ \ref{feynman}, respectively.} \label{zerlegung}
\end{figure}

   In order to investigate the total results for the Sachs form factors
in more detail, Fig.\ \ref{zerlegung} shows the individual
contributions of the different diagram sets for the fit up to
$Q^2_\textrm{max}=0.4$ GeV$^2$.
   The dotted lines represent the diagrams involving $\pi\Delta$ loops which seem
to adulterate the results for $Q^2\gtrsim0.3$ GeV$^2$.
   Similarly to the short-dashed lines involving $\pi N$ loops only, the $\Delta$ contributions
do not produce significant curvature.
   At larger values of $Q^2$, the strong linear $Q^2$ dependence of the dotted lines,
especially for $G_E^p$, $G_E^n$, and $G_M^p$, together with their
lack of curvature at $\mathcal{O}(q^3)$ cannot be compensated by the
strongly curved tree contributions of the vector mesons (long-dashed
lines).
   For this reason, the quantitative description of the data worsens
for larger values of $Q^2$ at the considered chiral order if the
$\Delta$ is included explicitly.
   In agreement with Ref.\ \cite{Schindler:2005ke}, the numerical contributions
resulting from the vector-meson loop diagrams, denoted by the dash-dotted
lines, turn out to be small.

\begin{figure}[t]
\begin{center}
\includegraphics[width=\textwidth]{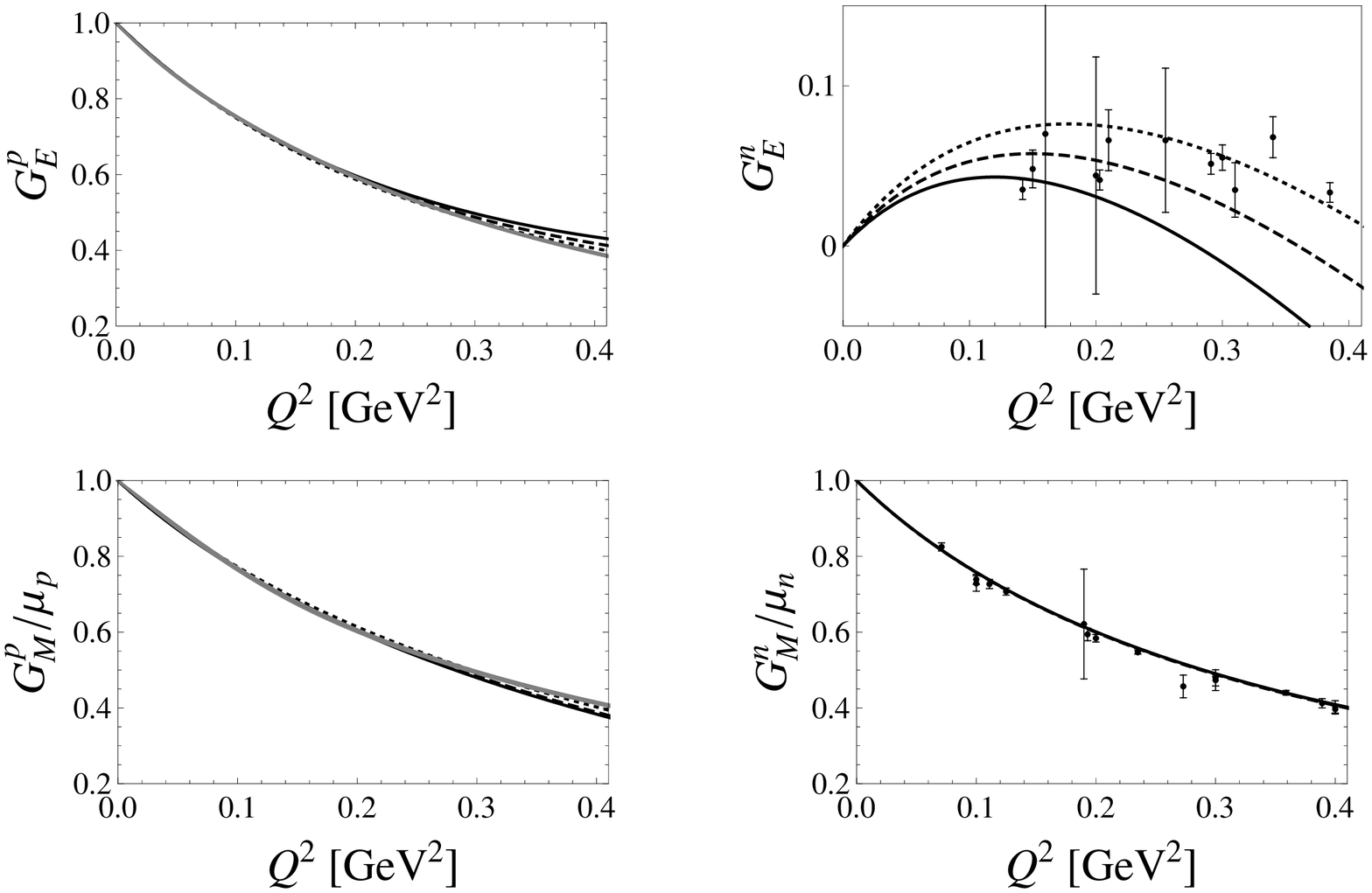}
\end{center}
\caption{Sachs form factors of the nucleon at $\mathcal{O}(q^3)$
including $\rho$ and $\omega$ fitted to different ranges of momentum
transfer $Q^2$. Full lines correspond to a fit up to
$Q^2_\textrm{max}=0.2$ $\textrm{GeV}^2$, dashed lines up to
$Q^2_\textrm{max}=0.3$ $\textrm{GeV}^2$, and dotted lines up to
$Q^2_\textrm{max}=0.4$ $\textrm{GeV}^2$, respectively.  The grey
bands represent empirical fits of the form factors to the measured
cross sections.} \label{VMFFQvar}
\end{figure}

    Figure \ref{VMFFQvar} displays the electromagnetic Sachs form
factors without explicit $\Delta$ resonance (see Table \ref{fit}).
   As opposed to Fig.\ \ref{DeltaFFQvar}, all fits describe the related data
remarkably well for all four Sachs form factors.
   Consequently, the consistent inclusion of $\rho$ and $\omega$ mesons provides
a satisfactory description of the electromagnetic form factors in
the momentum transfer region $0\leq Q^2\leq0.4$ $\textrm{GeV}^2$
already at third chiral order.
   In our calculation, the additional inclusion of the $\phi$ meson results
in only a small numerical adjustment of the parameters in Table
\ref{fit} and does not generate a visible improvement of the form
factor curves.

\begin{figure}[t]
\begin{center}
\includegraphics[width=\textwidth]{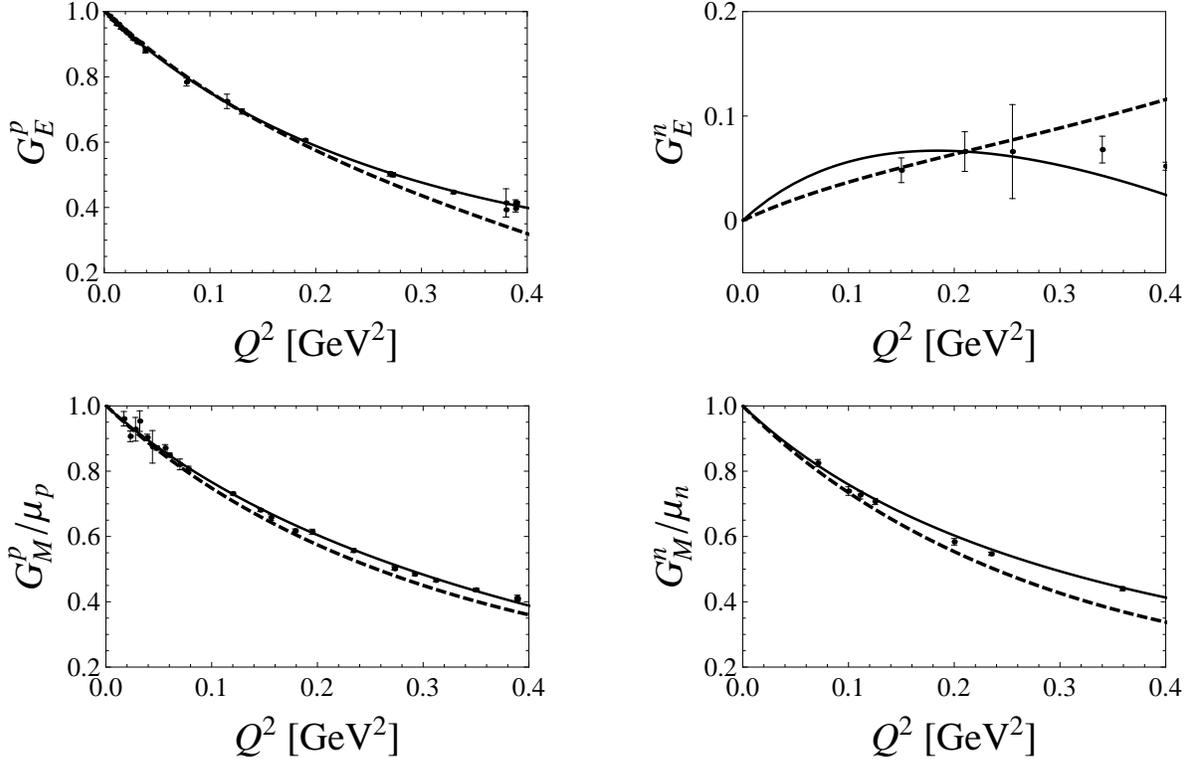}
\end{center}
\caption{Sachs form factors including vector mesons. The full lines
refer to the $\mathcal{O}(q^3)$ results of this work including
$\rho$ and $\omega$ mesons and the dashed lines to the
$\mathcal{O}(q^4)$ results of Ref.\ \cite{Schindler:2005ke}
including $\rho$, $\omega$, and $\phi$. The grey bands represent
empirical fits of the form factors to the measured cross sections.}
\label{vergleich}
\end{figure}

   In Ref.\ \cite{Schindler:2005ke}, the electromagnetic form factors of
the nucleon were calculated in Lorentz-invariant ChPT using the EOMS
renormalization scheme up to and including $\mathcal{O}(q^4)$.
   The vector mesons $\rho$, $\omega$, and $\phi$ were incorporated
explicitly using parameterization II of Ref. \cite{Ecker:1989yg}.
   The self-consistency relations for the $\rho$ meson couplings, as
discussed in Sec.\ III B, were not considered in the effective
Lagrangian.
   In order to discuss to what extent the self-consistent
inclusion of vector mesons influences the description of the form
factors, in Fig.\ \ref{vergleich} we display a comparison of our
calculation, including explicit vector mesons only, and that of
Ref.\ \cite{Schindler:2005ke}.
   Even though our calculation involves one vector-meson degree of freedom less,
namely the $\phi$ meson, and is only up to $\mathcal{O}(q^3)$, the
results for all Sachs form factors are slightly closer to
experimental data for $Q^2\gtrsim0.2$ $\textrm{GeV}^2$.
   In particular, our curve for the electric neutron form factor shows a better trend
for larger values of the momentum transfer.
   The improved description can be explained by the following two reasons.
On the one hand, the consistency relations lead to a parameterization
which features the $\rho\gamma$ and $\rho\pi\pi$ couplings at
leading order in distinction to the parameterization used in Ref.\
\cite{Schindler:2005ke} featuring such couplings only at
next-to-leading order.
   The re-ordering of terms changes the results favorably for larger values of $Q^2$.
   On the other hand, we fit our free coupling constants to the global trend of the
form factor curves instead of adjusting them to the electric and
magnetic radii and taking the remaining ones from results based on
dispersion relations.
   This approach allows for a better overall descriptions of the form factor curves.

\subsection{Estimate of higher-order effects}
\label{estimate}

\begin{figure}[t]
\begin{center}
\includegraphics[width=\textwidth]{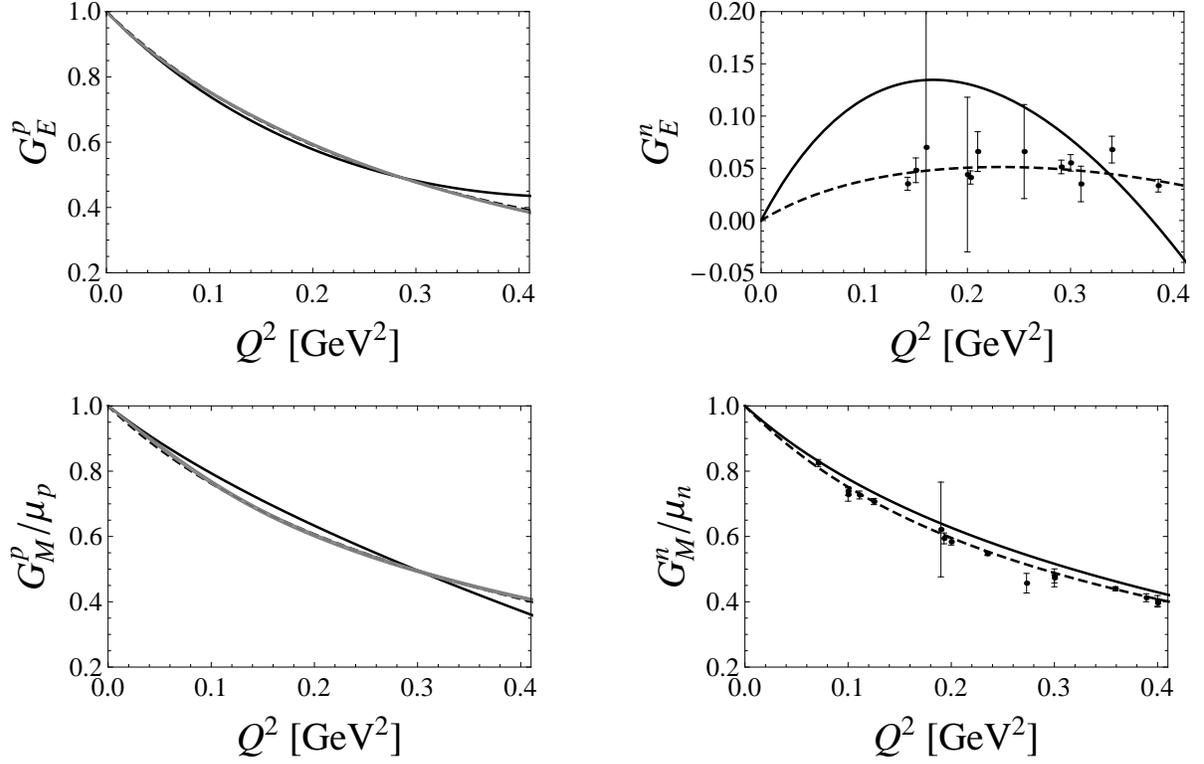}
\end{center}
\caption{Sachs form factors including vector mesons and the $\Delta$
resonance. The full lines refer to the $\mathcal{O}(q^3)$ results,
whereas the dashed lines are supplemented by additional higher-order
terms according to Eq.~(\ref{sachserror}). The grey bands represent
empirical fits of the form factors to the measured cross sections.}
\label{SachsErrorphys}
\end{figure}

\begin{figure}[t]
\begin{center}
\includegraphics[width=\textwidth]{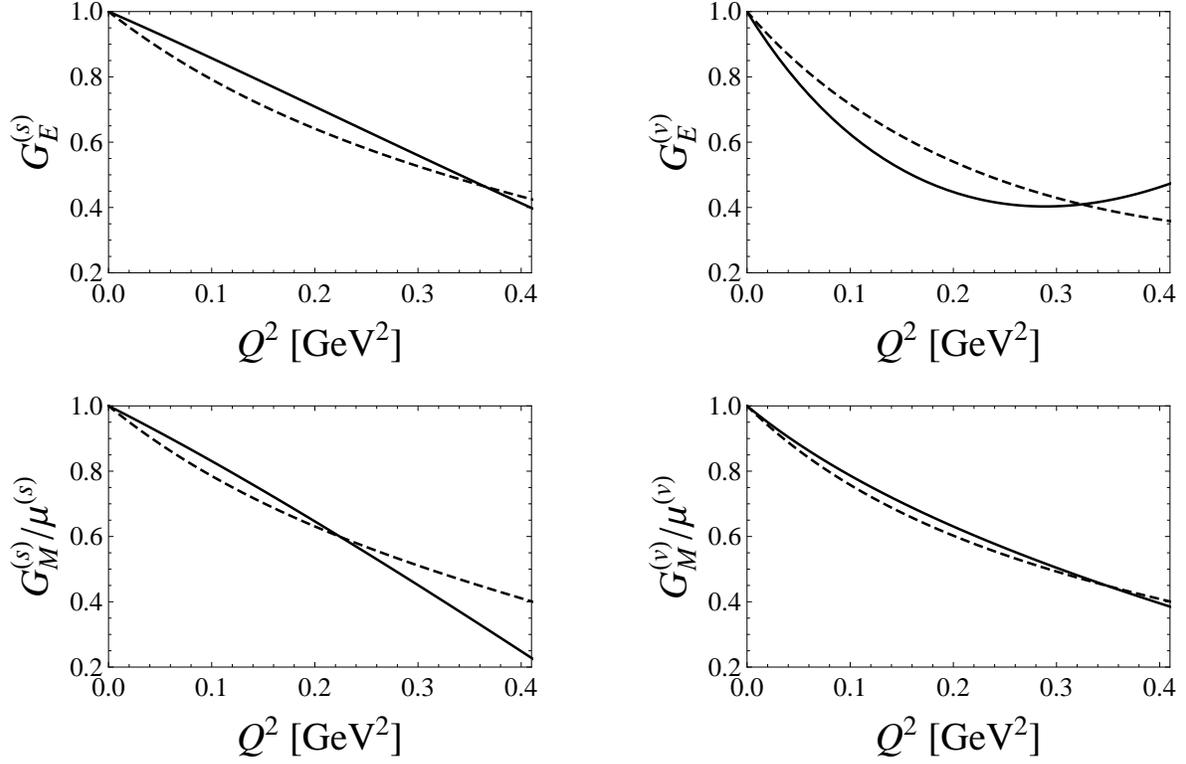}
\end{center}
\caption{Isocalar and isovector Sachs form factors including vector
mesons and the $\Delta$ resonance. The full lines refer to the
$\mathcal{O}(q^3)$ results, whereas the dashed lines are
supplemented by additional higher-order terms according
to Eq.~(\ref{sachserror}).} \label{SachsErroriso}
\end{figure}

    In order to estimate the uncertainties originating in the truncation
of the expansion at ${\cal O}(q^3)$, we add polynomials in $Q^2$ to the
Sachs form factors.
   The explicit form of the corresponding polynomial is motivated by an analysis
of the maximal powers of $Q^2$ in the isoscalar and isovector Dirac and Pauli
form factors, taking into account the available chiral structures at ${\mathcal O}(q^3)$
with adjustable LECs.
   We find that the neglected structures which would appear in an ${\mathcal O}(q^4)$
calculation can be parameterized as
\begin{align}
\Delta G_E^N&=\left(a_N \,Q^2- b_N\frac{Q^2}{4 m_p^2} \right)Q^2, &
\Delta G_M^N&=\left(a_N \,Q^2+b_N\right)Q^2,
\label{sachserror}
\end{align}
where $a_N$ and $b_N$ ($N=p,n$) denote unknown coefficients.
   The corrections to the electric Sachs form factors are purely of order $(Q^2)^2$
while those to the magnetic Sachs form factors already start at order $Q^2$.
   To investigate the influence of these higher-order contributions we add them to the
original expressions at ${\cal O}(q^3)$ and perform a new simultaneous fit of all
four "improved" Sachs form factors to the data.
   The results of this procedure are shown in Fig.\ \ref{SachsErrorphys}.
   Clearly, the inclusion of higher-order terms according to Eq.\ (\ref{sachserror})
changes the slopes of the Sachs form factors.
   The reason for this can be understood if one resolves
the physical Sachs form factors into their respective isoscalar and isovector
parts.
   Figure \ref{SachsErroriso} indicates that an ${\mathcal O}(q^3)$ calculation does
not generate enough curvature for both, the electric and magnetic, isoscalar form factors.
   To fit to the data over a wider range of momentum transfer in terms of minimizing
the total $\chi_{\text{red}}^2$ function, the respective slopes counterbalance the missing curvature.
   This overcompensation leads to a lower value for $\langle\left(r_M^p\right)^2\rangle$ and $\langle\left(r_M^n\right)^2\rangle$,
   as well as to a higher value for $|\langle\left(r_E^n\right)^2\rangle|$ than in the case with additional higher-order terms.
   A comparison between the radii obtained within the two approaches is shown in
Table \ref{radien2}.

\renewcommand{\arraystretch}{1.3}
\begin{table}
\begin{center}
\begin{tabular}{|c c c c|}
\hline  $\quad \langle\left(r_E^p\right)^2\rangle$ \quad&
\quad $\langle\left(r_E^n\right)^2\rangle$ \quad&\quad $\langle\left(r_M^p\right)^2\rangle$ \quad&\quad $\langle\left(r_M^n\right)^2\rangle$ \quad\\
\hline\hline
\quad $0.761$ \quad&\quad $-0.440$\quad&\quad $0.574$\quad&\quad $0.667$\quad\\
\quad $0.718$ \quad&\quad $-0.142$\quad&\quad $0.708$\quad&\quad $0.758$\quad\\
\hline
\end{tabular}
\end{center}
\caption{Mean square charge and magnetic radii of the nucleon obtained from the
form factor results including the $\Delta$ resonance. The first row
corresponds to the ${\mathcal O}(q^3)$ results, whereas the second row
also includes additional higher-order terms according to Eq.\ (\ref{sachserror}).
The mean square radii are given in units of fm$^{2}$. \label{radien2}}
\end{table}
\renewcommand{\arraystretch}{1}

\subsection{Pion mass dependence of the form factors}
\label{pion_mass_dependence}
   Nucleon electromagnetic form factors have been calculated in the framework of lattice QCD (for a recent overview,
see Refs.\
\cite{Yamazaki:2009zq,Hagler:2009ni,Bratt:2010jn,Collins:2011mk,Lin:2011sa}
and references therein). Several collaborations have reached pion
masses down to 270 MeV.
   Systematic extra\-polations to the physical value of the pion
mass and the $Q^2$ dependence are necessary to compare the lattice
form factor results to experimental data.
   Given the manifest Lorentz covariance of our results, they may
   provide
useful guidance for extrapolations of lattice simulations.
   In lattice calculations it is more convenient to work with isoscalar and
isovector form factors.
   Simulations of the isoscalar form factors are numerically expensive
since they involve the evaluation of disconnected quark loops.
   For the isovector form factors the disconnected quark loop contributions
cancel.

\begin{figure}[t]
\begin{center}
\includegraphics[width=\textwidth]{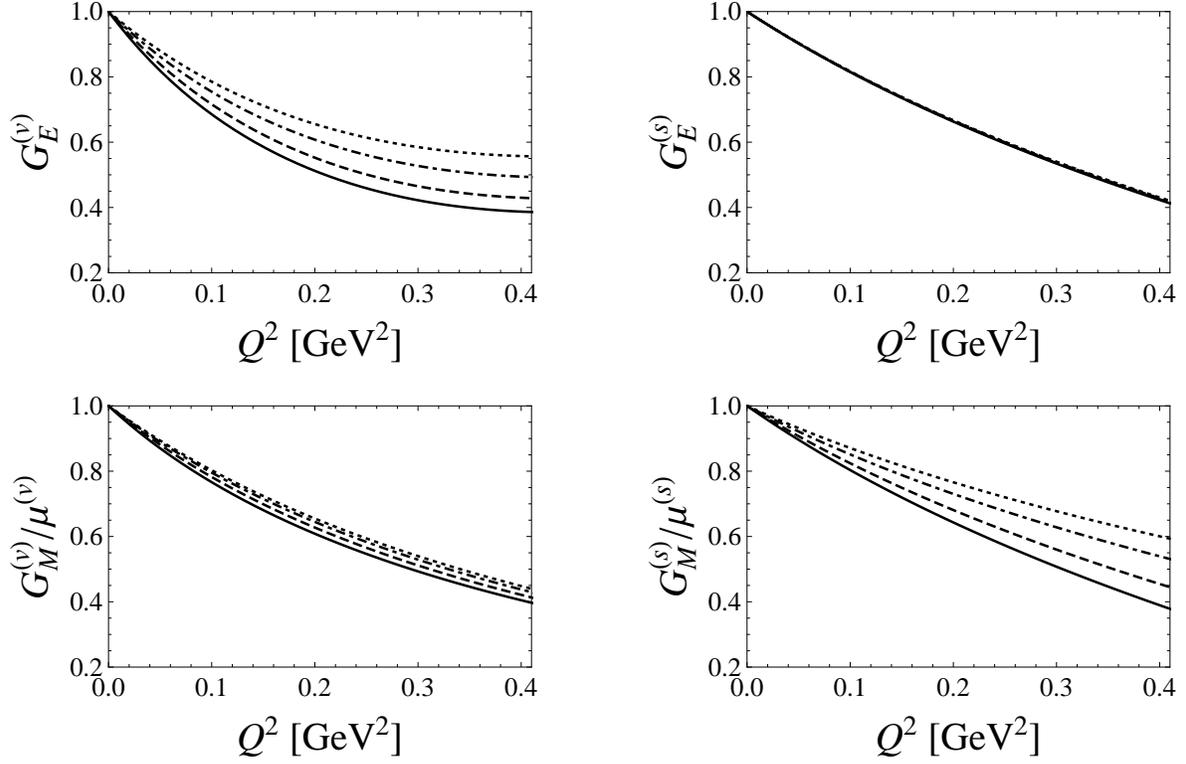}
\end{center}
\caption{Electric and magnetic Sachs form factors in the isovector
and isoscalar channel including $\rho$ and $\omega$ for different
values of the pion mass $M_\pi$. The full lines refer to $M_\pi=140$
MeV, the dashed lines to $M_\pi=200$ MeV, the dash-dotted lines to
$M_\pi=300$ MeV, and the dotted lines to $M_\pi=400$ MeV, respectively.}
\label{pionmassIso}
\end{figure}

\begin{figure}[t]
\begin{center}
\includegraphics[width=\textwidth]{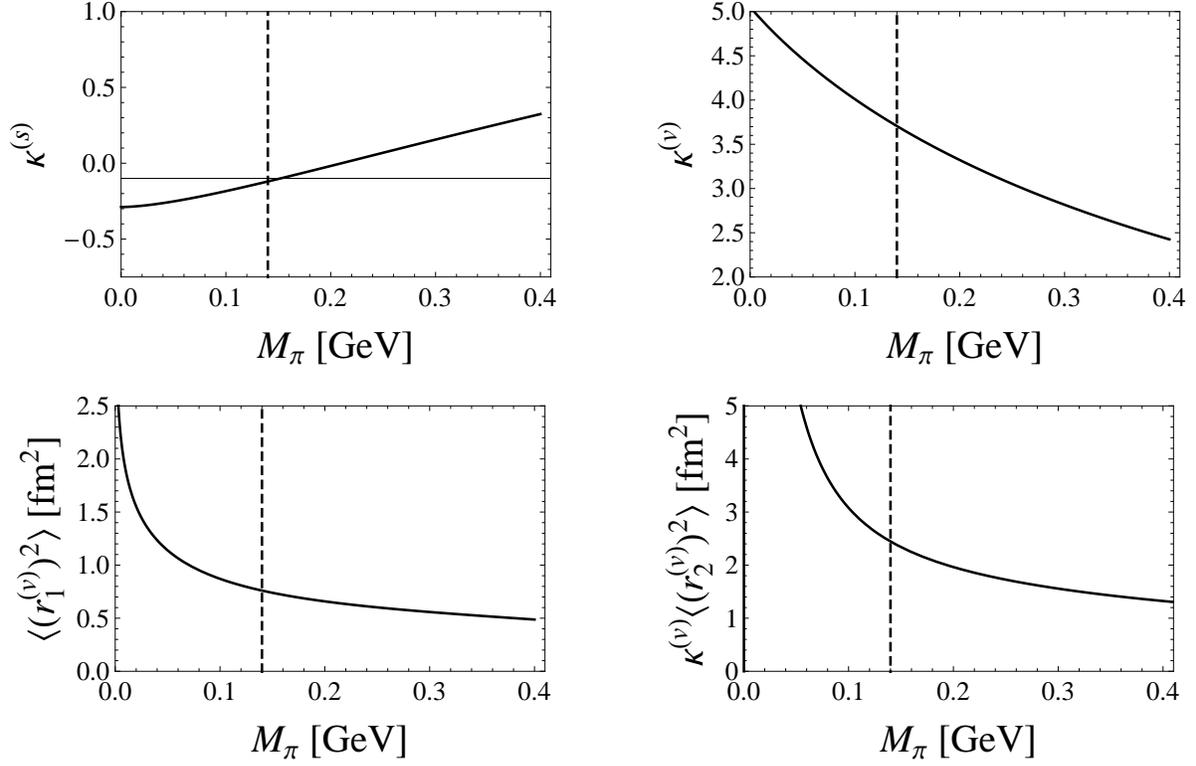}
\end{center}
\caption{Pion-mass dependence of the anomalous magnetic moment in
the isovector and isoscalar channels and of the mean square isovector Dirac and
Pauli radii; the vertical dashed line indicates the physical pion
mass.} \label{pionmassIso2}
\end{figure}

   In Fig.\ \ref{pionmassIso} we display the pion-mass dependence of the
Sachs form factors in the isovector and isoscalar channels.
   The quantities are derived from the results including $\rho$ and
$\omega$ mesons as explicit resonant degrees of freedom.
   We stress that, at the given order of our calculation,
the vector meson masses are independent of the pion mass (see, e.g., Refs.\
\cite{Djukanovic:2009zn,Feng:2010es,Feng:2011zk} for a discussion of the quark-mass dependence of the
$\rho$-meson mass).
   Extrapolations of lattice data of $G_E^{(v)}$ have been performed in
Refs.\ \cite{Lin:2010ne,Syritsyn:2009mx}.
   The results qualitatively agree with ours as the values of the form
factor fall off more slowly with increasing pion mass.
   In Ref.\ \cite{Hagler:2011zz} the isovector magnetic
moment is extrapolated to small pion masses using results from
heavy-baryon ChPT \cite{Hemmert:2002uh}.
   The pion-mass dependence and the value in the chiral limit are in
good agreement with our corresponding findings plotted in Fig.\ \ref{pionmassIso2}.

   In the present paper we fit the unknown LECs to experimental data at
the physical pion mass.
   Alternatively, they can be fitted to lattice simulation data at different
values of the pion mass resulting in a complete theoretical
prediction of the observables.
   Whether this is a useful approach largely depends upon the range of pion masses in
which the low-energy EFT is still applicable.
   The region of applicability depends on the calculational scheme and has yet to be
studied more thoroughly.

\section{Summary}

   We have calculated the electromagnetic form factors of the nucleon at
$\mathcal{O}(q^3)$ in manifestly Lorentz-invariant baryon chiral
perturbation theory including $\rho$ and $\omega$ mesons as well as
the $\Delta$ resonance.
   In terms of constraints and perturbative renormalizability, we have incorporated
the resonant degrees of freedom self-consistently into the EFT.
   To generate a systematic power counting we have applied the extended on-mass-shell
renormalization scheme.
   Two of the undetermined low-energy coupling constants have been adjusted to
the anomalous magnetic moments while the remaining six LECs have
been fitted simultaneously to the experimental data for different
ranges of momentum transfer up to $Q^2=0.4$ GeV$^2$.
   We found that the results incorporating vector mesons agree well with experimental
data in a momentum transfer region $0\leq Q^2\leq0.4$
$\textrm{GeV}^2$ while those also including the $\Delta$ describe
the form factors only up to $Q^2\approx0.3$ $\textrm{GeV}^2$.
   For larger values of $Q^2$, notably $G^n_E$ and $G^p_M$ disagree with the data.
   This is because the $\pi\Delta$ loop contributions at $\mathcal{O}(q^3)$ feature a
strong linear $Q^2$ dependence without sufficient curvature.
   Our results including $\rho$ and $\omega$ mesons at third chiral order agree at
least as well with experimental data as the previously performed
calculations of Refs.\ \cite{Kubis:2000zd,Fuchs:2003ir} at $\mathcal{O}(q^4)$.
   This improvement is a hint towards the importance of a proper consideration of
self-consistency relations among the couplings of the effective Lagrangian.
   The resulting form of the vector-mesonic Lagrangian gives rise to vector-meson
dominance at leading order and deviations thereof are pushed to higher orders.

\acknowledgments

   The authors thank J.\ Gegelia, H.\ B.\ Meyer, and M.~R.\ Schindler for
   useful
discussions.
   This work was supported by the Deutsche Forschungsgemeinschaft (SFB 443).

\end{document}